\documentclass{iau}
\usepackage{graphicx}

\title[IAUS 295.~~The intriguing life of massive galaxies] 
{A phenomenological approach to the evolution of galaxies}

\author[S. J. Lilly et al.]   
{Simon J. Lilly$^1$,
Yingjie Peng$^1$, Marcella Carollo$^1$
 \and Alvio Renzini$^2$}

\affiliation{$^1$Institute of Astronomy, ETH Zurich, 8093 Zurich, Switzerland 
\\[\affilskip]
$^2$INAF Ð Osservatorio Astronomico di Padova, vicolo dell'Osservatorio 5, I-
35122 Padova, Italy, and Department of Physics and Astronomy Galileo Galilei, Universita degli Studi di
Padova, via Marzolo 8, I-35131 Padova, Italy}

\pubyear{2013}
\volume{295}  
\pagerange{xx--xx}
\jname{The intriguing life of massive galaxies}
\editors{D. Thomas, A. Pasquali \& I. Ferreras, eds.}
\begin{document}

\maketitle

\begin{abstract}
Increasingly good statistical data on the galaxy population at high and low redshift enable the development of new phenomenological approaches to galaxy evolution based on application of the simplest continuity equations.  This has given new insights into the different ways in which star-formation in galaxies is quenched, the role of merging in the population, and in to the control of star-formation in star-forming galaxies and the links with chemical evolution.  The continuity approach provides a self-consistent view of the evolving population and exposes linkages between different aspects of galaxy evolution.

\end{abstract}

\firstsection 
\section{Introduction: the continuity approach to galaxy evolution}

Large surveys of the extragalactic Universe, both nearby and at high redshifts, are producing high quality statistical data on the galaxy population over a wide range of epochs.  These allow the possibility of a new phenomenological approach to galaxy evolution that is built around the simple continuity equation(s) that link the galaxy population at different epochs.   In this approach, we first identify, both locally and at high redshift, a few underlying simplicities, or symmetries, exhibited by the galaxy population and then explore analytically the implications of these via the continuity equations.  

This approach is in a sense reversed from the usual theoretical approach of the semi-analytic models, which start with the hierarchical build-up of dark matter haloes, within which baryons evolve subject to a large number of physical processes that are described by analytic formulae.  The models end up involving a large number of overt free parameters (at least 30) plus many additional assumptions that are buried in the implementation.  State of the art semi-analytic models are quite successful in reproducing the galaxy population, but the inevitable complexity of the models means that there is considerable uncertainty as to the uniqueness of any particular implementation, and agreement with observations is never perfect. 

Our continuity approach steps back from physical inputs (apart from the most basic continuity equations) and focuses directly on what the {\it data} appear to require.  This approach produces phenomenological descriptions of the evolution of galaxies that may help constrain more physically motivated models.  It can also identify hidden connections between different aspects of galaxy evolution that appear at first sight to be disjoint.   At the very least, this approach should produce a simple self-consistent framework for interpreting different data, provide a common language for discussion, and allow well-posed questions to guide future observational programs. 

We here review our recent work in this area, and discuss some of the insights that have been gained in a series of recent papers.   In Section 2 (based on Peng et al 2010, 2012, hereafter P10 and P12), we focus on the inter-relationships between star-formation, stellar mass and environment and the ``quenching'' of star-formation in galaxies. In Section 3 (based on Peng et al, in preparation) we briefly discuss the merging of galaxies which is also discussed in another contribution to these Proceedings. Finally, in Section 4, which is based on Lilly et al (2013, hereafter L13), we examine the control of star-formation in star-forming galaxies, and the connections between metallicity, the cosmic star-formation history of galaxies and the efficiency with which dark-matter haloes produce stars.    

\section{The quenching of galaxies}

Most star-forming galaxies have an SFR that is well-correlated with their existing stellar mass, producing a tight ``Main Sequence'' (MS) on which the specific SFR (sSFR) depends only weakly on stellar mass, $sSFR \propto m^{\beta}$ with $\beta \sim -0.1$, with a scatter of order 0.3 dex.   A few percent of galaxies have significantly elevated sSFR, but these contribute only about 10\% of the overall star-formation over a range of redshifts $0 < z < 2$ (Rodighiero et al 2011, Sargent et al 2012).  Locally, these star-bursts are associated with major mergers of galaxies, and the same may be true at high redshifts also.   Finally, there is a substantial population of galaxies with much reduced sSFR.  These ``quenched" galaxies are not forming a significant number of stars, i.e. $sSFR^{-1} >> \tau_H$.   The physical process that turns off star-formation in galaxies is not well understood.  Plausible ideas include the action of AGN or extreme merger-induced star-bursts, the inability of some haloes to deliver cool gas to the central galaxy, plus a whole host of ways in which satellite galaxies may interact with the environment.

\subsection{Key observational axioms}

Our analysis of quenching is based two key observational facts, which we use as axioms:

\begin{itemize}

\item {\bf Separability of the red fraction}:  As shown in P10, the SDSS red fraction $f_{red}(m,\rho)$ is separable in mass and environment (``environment" is here a nearest neighbour measure of local density, but could equally well be a central-satellite division etc.).  This means that the fraction of ``surviving'' blue galaxies can be written as the product of two functions, one of mass but not environment, and the other of environment but not mass.  
\begin{equation}\label{separability}
f_{blue}(m,\rho) = (1-\epsilon_m(m)) \times (1-\epsilon_{\rho}(\rho))
\end{equation}
Separability was already implicit in one of the fitting formulae for the SDSS $f_{red}(m,\rho)$ given by Baldry et al (2007). We argued in P10 that this separability suggests that there are two ``probabilistic'' channels for quenching galaxies: one we call ``mass-quenching'', which is linked to the mass of galaxies but not environment, and the other, ``environment-quenching'', determined by the environment of a galaxy but not by its stellar mass.   Of course, galaxies are unlikely to evolve probabilistically, and so there are presumably hidden parameters at work, but these must be independent of environment for mass-quenching, and vice versa.  

\item {\bf The shape of the mass function of star-forming galaxies stays the same back to at least $z \sim 2$ and probably back to $z \sim 4$. }:  Both the characteristic mass M* and the faint-end slope $\alpha_S$ of the Schechter mass function of star-forming galaxies show very little evolution between SDSS locally and the high redshift population. In contrast, the density normalization $\phi*$ increases by a factor of about 3 since $z \sim 2$, and by 20 since $z \sim 4$.   This ''vertical'' evolution in $\phi_{SF}(m)$ is quite different from the ''horizontal'' evolution of the mass-function of dark matter haloes, which changes primarily in $M*$ with more or less constant $\phi*$. The surprising constancy of $M*$ was first highlighted by Bell et al (2005) to $z \sim 1$, was evident in Gonzales et al (2010) to $z \sim 4$ using photo-$z$ to $z \sim 4$, and is firmly established in COSMOS with both photo-$z$ and spec-$z$ (Ilbert et al 2010, Pozzetti et al 2010, Ilbert et al 2013).  

\end{itemize}

Other important observations that simplify the analytic treatment, but are not really required per se, are (i) the small dispersion in sSFR on the Main Sequence, (ii) the weak dependence of the Main Sequence sSFR on mass, $\beta \sim 0$, and (iii) the evidence that the environment term $\epsilon_{\rho}(\rho)$ is more or less constant back to $ z \sim 0.8$ (P10).   It is also convenient to assume that quenching is instantaneous and one-way.  The characteristic sSFR of the MS gives the rate at which galaxies increase their logarithmic mass, and the evolving sSFR(t) therefore ``sets the clock'' of the evolving population but does not affect the outcome. 

These most basic observational results are of course inter-linked.  The dramatic increase in sSFR to high redshift implies that a given galaxy that stays on the Main Sequence will have increased its mass by a factor of about 50 since $z \sim 2$. This makes the constancy of $M*$ and $\alpha_S$ all the more striking, and emphasizes the importance of fluid-like continuity equations to track the evolving galaxy population.

\subsection{Environment-quenching as satellite-quenching}

In P12, it was shown that the environmental effects in SDSS that we highlighted in P10 are confined to satellite galaxies. Unlike the satellites, the properties of central galaxies do not depend on $\rho$ (at fixed mass).  The satellite-quenching efficiency, $\epsilon_{sat}$, is the probability that a blue star-forming galaxy is quenched because it becomes the satellite of another galaxy.  This is empirically independent of stellar mass in SDSS (as in van den Bosch et al 2008) with a mean value of $\epsilon_{sat} \sim 0.4$.  As predicted from P10, $\epsilon_{sat}$ increases with the local density $\rho$ which may be taken as an (imperfect) proxy for the location within a group/cluster. It is largely independent of group mass (at fixed $\rho$).   In other words, the environment-quenching of P10 is simply ''satellite-quenching''.  Similar results are also evident in the detailed study of 170 low redshift ZENS groups (Carollo et al, in preparation).

Satellite-quenching is independent of stellar mass, and so it cannot change the shape of the mass-function of star-forming galaxies. Instead, it will produce a mass-function of satellite-quenched passive galaxies that has the same $M*$ and $\alpha_S$ as the star-forming population, but a relative $\phi*_{EQ}/\phi*_{SF}$ that increases with the local density.

\subsection{Mass-quenching}

Mass-quenching acts on all galaxies, independent of environment, i.e. on both centrals and satellites. Since only mass-quenching is linked to galaxy mass, it must be this process alone that controls the shape of the mass function of the (surviving) star-forming galaxy population.  Indeed, ''mass-quenching'' is usefully thought of as {\it that} process which controls the mass function of star-forming galaxies, and the star-forming mass-function therefore gives us as clear a view of the action of mass-quenching as the more traditional study of the red fraction.  

The observed constancy of $M*$ for the star-forming population places a strong requirement on the form of mass-quenching. This can be expressed either as a quenching rate $\lambda_m$ (the chance of a given galaxy to be mass-quenched per unit time) or, equivalently, as a survival probability for a given galaxy to reach a particular mass $m$ without being quenched, $P(m)$.   Both versions include a constant $\mu$ that is the inverse of the (time-independent) $M*$, i.e. $\mu = M*^{-1}$.  It is easy to show analytically from the continuity equation that $\lambda_m(m,t)$ and $P(m)$ must have the following form
\begin{equation}\label{mass-quenching}
\lambda_m = \mu \times SFR  \Longleftrightarrow P(m) = exp(-\mu m)
\end{equation}

\subsection{The origin of the Schechter function(s)}

The action of the mass-quenching described by Equation (\ref{mass-quenching}) not only maintains the M* of the (single) Schechter function of star-forming galaxies (as it must, by design) but it will also {\it establish} a precise Schechter shape starting from a more general power-law mass-function of the right slope.  The mass function of the mass-quenched passive galaxies also has a particularly simple form: it too is a single Schechter function, with exactly the same M* as the star-forming population, but with a faint end slope that differs by $\Delta\alpha_S = (1+\beta) \sim 1$.  Given that $M*$ for the star-forming population is observed to be constant, the M* of the (mass-quenched) passive galaxies will also evolve with constant M* as the population builds up over time.   The continuity equation also requires that the $\phi*$ of the two populations should increase in step, with $\phi*_{MQ}/\phi*_{SF} = -(1+\alpha_{S,SF})^{-1} \sim 2.5$.   

The superposition of these different Schechter functions for the mass-quenched and environment-quenched passive galaxies produces a double Schechter function for the overall passive population. The second component depends on environment and will be absent in centrals.   Adding the star-forming population, the overall population will also be a double Schechter function. Interestingly, the strength of the two components in this overall double Schechter function does not depend on the amount of environment quenching and is given by the same relation as before, $\phi*_{1+\alpha_S}/\phi*_{\alpha_S} = -(1+\alpha_{S})^{-1} \sim 2.5$

These quantitative relations between the Schechter parameters are seen, to impressive precision, within the SDSS sample as a function of environment (P10), and for centrals/satellites (P12, see Figure 1).  Not least, the characteristic inflection of the double Schechter function due to the superposition of the red and blue populations, is clearly visible in the SDSS, GAMA and zCOSMOS mass-functions (Pozzetti et al 2010, P10, P12, Baldry et al 2012).

\begin{figure}[]
 \vspace*{0 cm}
\begin{center}
\includegraphics[angle=0,origin=c,width=5.3in]{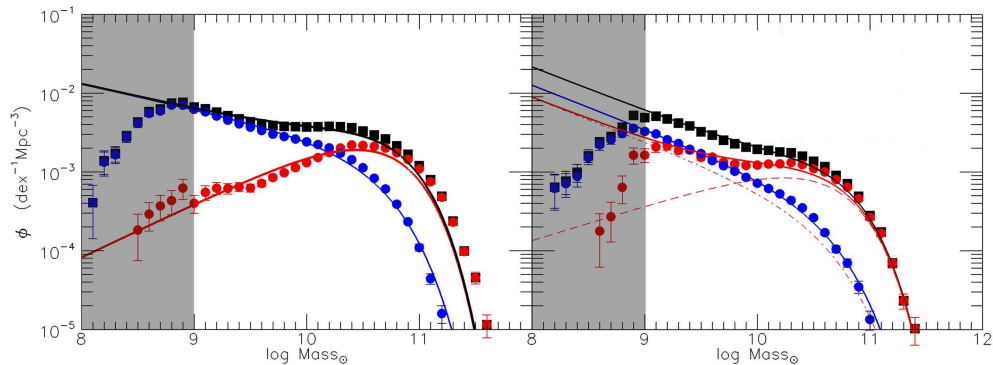} 
\vspace*{-0.5 cm}
\caption{The mass functions of SDSS centrals (left panel) and satellites (right panel) that are observed in the SDSS, split into blue and red populations (as indicated), plus the overall population (in black).  In each case, the mass functions of the red (plus overall populations) that are {\it predicted} on the basis of the {\it observed} blue population using the P10/P12 formalism are shown as continuous lines (split in the case of satellites into the mass-quenched and satellite-quenched components as the dashed lines). These mass-functions provide an excellent representation of the data.  The excess of high mass red central galaxies is interpreted as a modest amount of mass-increase through merging after quenching. Figure is taken from P12. }
\label{mf}
\end{center}
\end{figure}

\subsection{Interpretation of mass-quenching}

We stress that the $\lambda$ form of Equation (\ref{mass-quenching}) does not necessarily require that mass-quenching is physically caused by star-formation (or by stellar mass, as sometimes assumed), only that the quenching rate must {\it mimic} the mass- {\it and} time- dependencies of the SFR in MS galaxies.  

Given the known observational similarities between SFR and black hole accretion in galaxies (Silverman 2009), quenching due to AGN activity could satisfy this requirement.  Furthermore, given the close coupling between $m_{star}$ and $m_{halo}$ for central galaxies, the second, $P(m)$, form could conceivably be connected to a limit in the $m_{halo}$ that can sustain star-formation in its central galaxy.  Note, though, that the $M*$ of central and satellite star-forming galaxies are essentially identical in SDSS (P12) and that the satellite $M*$ is constant over $12.5 < {\rm log} (M_{halo}/M_{\odot}) < 15$ (see P12), so the mass-quenching process must operate identically in centrals and in satellites over a very wide range of halo mass.  Also, the observation that the $f_{red}$ of centrals is correlated with halo mass, at fixed stellar mass (e.g. Woo et al 2012), does not necessarily imply a causal link between quenching and halo mass as it can arise by the continued increase of halo mass, but not of stellar mass, after a galaxy has been quenched (see P12).  Finally, we stress that mass-quenching need not be physically caused by stellar mass either - indeed it is clear that morphology (e.g. central mass density, concentration or Sersic parameter) is a better ``predictor'' of the sSFR than $m_{star}$, even though causal basis of this link is unclear.  The only thing we know for sure is that mass-quenching must satisfy Equation (\ref{mass-quenching}).

\subsection{Quenching at higher redshifts}

Densely sampled spectroscopic surveys like zCOSMOS and DEEP allow the construction of group catalogues similar to those in SDSS at redshifts out to $z \sim 1$.  The $\epsilon_{sat}$ appears to be still independent of stellar mass at $z \sim 0.8$ (Knobel et al 2013) with a mean value of $\epsilon_{sat} \sim 0.5$ close to that seen in SDSS.  The mass functions of star-forming and passive centrals and satellites at high $z$ also follow the interrelations expected from the simple P10/P12 model.  The form of the mass- and environment- quenching efficiencies look similar, the role of satellites in driving the latter appears to be the same, and the form of $\epsilon_{sat}(\rho)$, all seem to be consistent at $z \sim 0.8$ with what is seen locally (Kovac et al, in preparation). We conclude that the action of the environment on satellite galaxies appears to be rather similar when the Universe was a half its current age, as today.  This provides another interesting constraint on the physical processes involved.

\subsection{Overall evolution of the population}

The action of the simple quenching laws is well seen in a movie that is available from http://iopscience.iop.org/0004-637X/721/1/193/fulltext/apj351504f13.mov.  

At early times, almost all galaxies are star-forming and the population evolves primarily by increasing in mass at a rate given by the rsSFR of the population.  We call this ``horizontal'' evolution of $\phi(m)$ as Phase 1 of the evolution.   As galaxies approach what will become the characteristic $M*$, mass-quenching becomes increasingly important and starts to produce massive quenched galaxies.  The star-forming $\phi(m)$ stabilizes as a single Schechter function (if it is not one already) and thereafter increases in $\phi*$ with constant $M*_{SF}$.  The mass-quenched passive population also increases in $\phi*$ at the same rate (with the same constant $M*$ but different $\alpha_S$).  This ``vertical'' evolution of $\phi(m)$ is Phase 2.  Finally, the quenching of satellite galaxies becomes important, producing the second component of passive galaxies which increasingly comes to dominate at lower masses.  We call this Phase 3 although it proceeds in parallel with Phase 2.  The transition between Phase 1 and 2 is marked by the establishment of constant $M*_{SF}$ and the appearance of the first massive passive galaxies, and represents a divergence in evolution between the overall galaxy and halo mass functions.  This probably occurs sometime around $z \sim 3-4$ and quite possibly at different times in different large scale environments.

An important point is that the environment- (or satellite-) quenched population of passive galaxies, which dominates at lower galaxy masses, emerges later compared to the population of mass-quenched galaxies, which dominate the higher masses around $M*$. The mass-quenching rate was twenty times higher at $z \sim 2$ than now, whereas the environment-quenching rate has evolved more gradually as groups and clusters were assembled.  This difference accounts for a correlation between mean stellar age and stellar mass for today's passive population, as well as the over-abundance of $\alpha$-elements in the more massive galaxies (see P10 for details).   It also naturally accounts for the apparent mass-{\it dependent} differential build up of the passive population (see Ilbert et al 2013), even though both channels, individually, build up the passive population {\it independently} of mass. These observational effects are both sometimes taken as a signature of a ``down-sizing'' process, but this is rather misleading since we are dealing with two distinct processes proceeding at different rates.

The continuity approach also clarifies two other widespread misconceptions:  It is clear from the mass-functions that there is today a threshold mass above which most galaxies are passive and below which most are still actively forming stars. It is often assumed that, in order to produce today's massive passive galaxies in the early Universe, this threshold must have been shifted to higher masses at earlier times (at variance with the observed constant M*).  Alternatively, it is assumed that massive passive galaxies were generally produced at lower masses and then moved to higher masses via widespread dry-merging.  The continuity analysis emphasizes that both star-forming and passive populations have very similar values of M*, and the different shapes of $\phi(m)$ come about purely from the different $\alpha_S$ of the two populations, that are themselves a result of the particular, required, form of the mass-quenching process. The observed (constant-shape) star-forming population can make all except the most massive of today's passive population directly.

Finally, the continuity formalism makes a clear prediction for the mass function of any set of objects that are seen at some stage in the process of being quenched.  Even if they are still actively forming stars, these transition systems should have the same $M*$ and $\alpha_S$ of the {\it passive} population, but a $\phi*$ given by the $\phi*$ of the {\it star-forming} population multiplied by $\tau_{T} \times sSFR$, where $\tau_T$ is the duration of the relevant phase of quenching, see Equation (28) of P10.  

\section{Merging of galaxies}

The merging of galaxies enters into the picture in at least two ways. This is discussed in Peng et al (in preparation), and also in the paper by Peng et al in these Proceedings.  Here we very briefly summarize the results, for completeness:

\begin{itemize}

\item {\bf Mass-addition after-quenching}:  Substantial addition of stellar mass to passive galaxies after they have been quenched would perturb the quantitative relations between the Schechter functions of star-forming and passive galaxies outlined above.  This limits the average mass increase for passive galaxies to be quite
modest, i.e. $<25\%$ for passive galaxies generally, and $<35\%$ for passive centrals.  

\item{\bf Maintaining constant $\alpha_S$}: If the slope of the sSFR-mass relation is negative, then low mass galaxies will increase in logarithmic mass faster than more massive ones, potentially producing a steepening of the mass function, which is not seen.  The destruction of low mass galaxies through merging provides an obvious way of countering this.  If we define a {\it specific merger mass rate} ($sMMR$) that is the mass increase of a galaxy  through merging with smaller galaxies (as opposed to the in situ star-formation described by the rsSFR), then it turns out that an sMMR $\sim$ 0.1 sSFR will compensate for the steepening effect of associated with $\beta \sim -0.1$.  An obvious physical explanation for the apparent link between merging and cosmic sSFR is that both are being driven by the specific growth rate of dark matter haloes, as discussed further in Section \ref{bathtub} below.  The fact that both the merger rate of galaxies and the specific star-formation rate follow the growth of haloes produces the ``symmetry'' or "simplicity" of constant $\alpha_S$ in the overall population.

\item{\bf Merger quenching}: The introduction of ``merger-quenching'' as a potential quenching channel in P10 was one of the few ad hoc aspects of that paper, which we have now clarified:  If $\epsilon_{\rho}(\rho)$ is to remain more or less constant (which was not imposed as a condition in P10) then the $\kappa_-$ term in Equation (18) of P10 must have a specific form: $\kappa_-(\rho) =\epsilon_{\rho}(\rho)\times sSFR$.  This link with $\epsilon_{\rho}(\rho)$ makes it attractive to view this term as part of the general "environment-quenching" process, which may or may not be linked to merging {\it per se}, rather than as a distinct third quenching channel.  

\end{itemize}

\section{The evolution of cosmic star-formation rate, haloes and metallicity}\label{bathtub}

A similar phenomenological approach to star-forming galaxies also reveals linkages between different aspects of galaxy evolution - namely the epoch dependence of the characteristic sSFR of MS galaxies, the chemical evolution of galaxies, and the varying efficiencies with which dark matter haloes have converted gas into stars (see Lilly et al 2013, hereafter L13, for details).

\begin{figure}
 \vspace*{-2.7 cm}
\begin{center}
\includegraphics[angle=0,origin=c,width=5in]{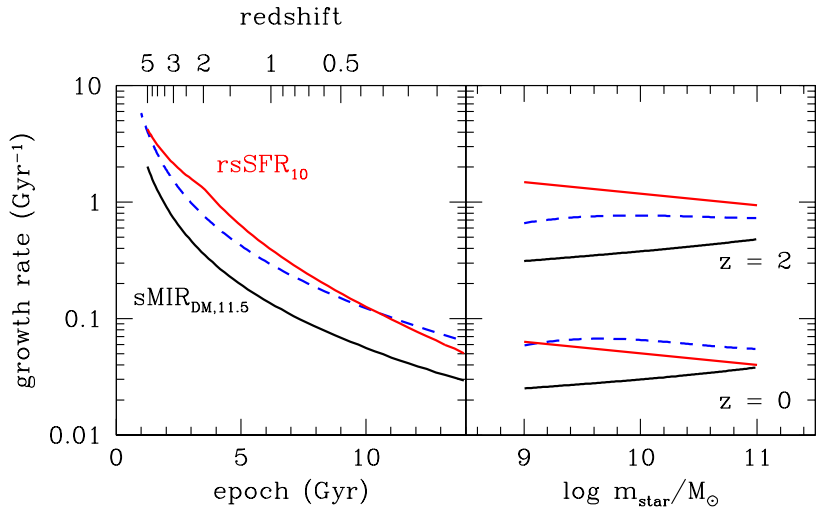} 
\vspace*{-2.5 cm}
\caption{Comparison of the specific growth rate of dark matter haloes, sMIR$_{DM}$ (in black) from simulations, and of the stellar populations of galaxies, given by the observed rsSFR (in red), as a function of epoch for two particular masses ($10^{10}$ and $10^{11.5}$ M$_{\odot}$ respectively in the left panel, and as a function of mass at $z = 0$ and $z = 2$ in the right panel.  These curves show the systematic offset of the sMIR$_{DM}$ to higher values (shorter doubling times) and the reversal of the mass-dependence of the $sMIR$ relative to the rsSFR.  The blue dashed curves show the boosting of the rsSFR relative to the $sMIR_{DM}$ expected from the $f_{star}(m)$ dependence - see text.}
\label{sSFR}
\end{center}
\end{figure}

We again start with a striking fact about the galaxy population:  The increase in the sSFR of Main Sequence galaxies back to $z \sim 2$ (Noeske 2007, Daddi 2007, Elbaz 2007 and Panella et al 2009) is very similar to the change in the specific mass increase rate of dark matter haloes (sMIR$_{DM}$) over the same interval (see Neistein \& Dekel 2009).  Above $z \sim 2$ the situation becomes less clear:  Initial evidence for a more or less constant $sSFR(z)$ (Perez-Gonzalez et al 2009) has been disputed (Stark et al 2012, Schaerer et al 2012).  Note that we must use a ``reduced'' sSFR, i.e. ($rsSFR = (1-R) \times sSFR$) to give the mass-doubling timescale of the long-lived stellar population (see L13). 
These similarities are shown in Figure (\ref{sSFR}). There are, however, important differences in detail: note that the rsSFR for typical mass galaxies is is consistently significantly higher, especially at high redshifts, implying a shorter doubling time, and that the weak mass-dependence is reversed, with $\beta_{DM} \sim +0.1$ compared to $\beta \sim -0.1$ for star-formation.

\subsection{Gas-regulated evolution of galaxies}

A very simple model for galaxies can be constructed (see also Bouche et al 2010) in which the star-formation rate in a galaxy is regulated by the gas content of the galaxy, via some star-formation efficiency $\epsilon = SFR / m_{gas}$, so that $\epsilon^{-1}$ is the gas depletion time-scale $ \tau_{gas}^{-1}$.   This implies that the $sSFR$ and the gas ratio $\mu = m_{gas}/m_{star}$ are linked by
$sSFR = \epsilon\mu$.
The star-formation may drive a wind $\lambda \times SFR$, where $\lambda$ is the mass-loading. The inflow can be parameterized to be a fraction $f_{gal}$ of the baryonic inflow into the halo.   If $f_{gal}$ and the internal parameters $\epsilon$ and $\lambda$ are all independent of time, then this model has the attractive feature of quickly setting and maintaining the $rsSFR$ to be exactly equal to the $sMIR_{DM}$ of the halo.  It does so by forming a constant fraction $f_{star}$ of the inflowing baryons into stars - the actual fraction depending on $\lambda$ and the $(\epsilon \times sSFR)$ product.   If these parameters of the regulator change with time, either directly or because they are functions of the stellar mass, then this equality will be perturbed, as discussed below.  

\subsection {Chemical evolution}

For as long as the gas consumption timescale is shorter than the timescales on which $\lambda$ and $\mu$ ($= \epsilon \times sSFR)$ are changing, gas can be considered to be continuously flowing through the system. The metallicity of the gas in this quasi-steady-state flow is given by
\begin{equation}\label{Z1}
Z = Z_0 + {y_R \over (1+\lambda(1-R)^{-1} + \epsilon^{-1}(sSFR + (1-R)^{-1}d{\rm ln}\mu/dt)}
\end{equation}
where $Z_0$ is the metallicity of the inflowing gas (assumed to be small) and $y_R$ is the yield of metals per unit mass locked up in long-lived stars.  The metallicity of the system is to a large degree set ``instantaneously'' by the regulator. The only knowledge of the previous history is in the (small) $\epsilon^{-1}d{\rm ln}\mu/dt$ term.  As would be expected in a continuous flow model, Equation (\ref{Z1}) can be written simply in terms of the fraction $f_{star}$ of incoming baryons that are transformed into stars, 
\begin{equation}\label{Z2}
Z = Z_0 + y_R \times f_{star},
\end{equation}
without knowledge of the quantities $\lambda$ and $\mu = \epsilon^{-1}sSFR$ which actually set the value of $f_{star}$.  These two Equations (\ref{Z1} and \ref{Z2}) highlight several inter-connections between different aspects of galaxy evolution.

\subsection{The existence of SFR as a second parameter in the mass-metallicity relation} 

Equation (\ref{Z1}) requires that the SFR should enter as a second parameter in the mass-metallicity relation, as has been claimed observationally in several papers (Ellison \& Kewley 2008, Manucci et al 2010, Lara-Lopez et al 2010, Yates et al 2012).   Furthermore, because the metallicity in the regulator is set ``instantaneously'', the $Z(m,SFR)$ relation should only evolve with time in so far as the parameters $\epsilon$ and $\lambda$ themselves change with time.  Equation (\ref{Z1}) provides a natural explanation for a more or less epoch-independent ``fundamental'' $Z(m,SFR)$ relation (e.g. Manucci et al 2010).

The derivation of gas-phase metallicities from emission line data is fraught with uncertainties (e.g. Kewley \& Dopita 2002) and there has been a lot of debate as to the form of the $Z(m,SFR)$ relation, even at low redshift (see e.g. Yates et al 2011).  With this significant caveat in mind, we can fit Equation (\ref{Z1}) to the SDSS $Z(m,SFR)$ data tabulated by Mannucci et al (2010) and examine the recovered $\epsilon(m)$ and $\lambda(m)$.  For simplicity, $y_R$ and $Z_0$ are assumed to be independent of mass.  Intriguingly, the $\epsilon(m)$ and $\lambda(m)$ are not unreasonable from independent astrophysical arguments: $\epsilon(m)$ implies a (total = atomic+molecular) gas depletion timescale at $m \sim 10^{10}$M$_{\odot}$ of 2-3 Gyr (c.f. Genzel et al 2010, Tacconni et al 2012), with a weak inverse mass dependence, while the nominal wind-loading factor at this mass is $0.3 < \lambda < 0.5$ with a quite strong inverse dependence on stellar mass, $m^{-0.8}$.  

\begin{figure}
 \vspace*{0 cm}
\begin{center}
\includegraphics[angle=270,origin=c,width=4in]{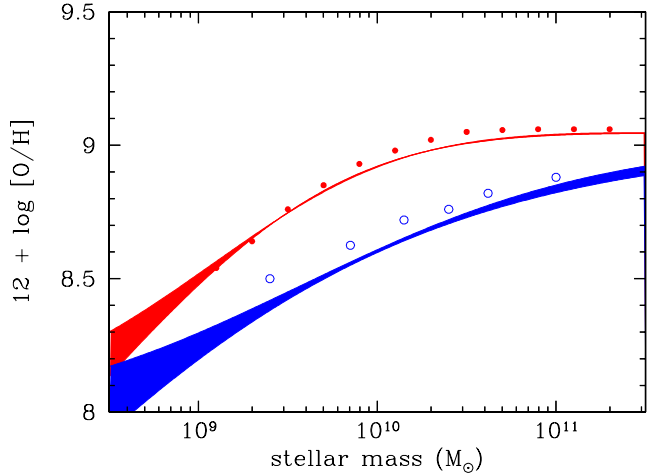} 
\vspace*{-2. cm}
\caption{The mean mass-metallicity relation at $z \sim 2$ (blue) and $z \sim 0$ (red). The data is from Mannucci et al (2010) and Erb et al (2004) - the latter adjusted by 0.3 dex, see L13 for details.  The simple model described in the text and given by Equation (\ref{Z1}) qualitatively reproduces the observed evolution, which is primarily driven by the much higher sSFR at high redshift.   The Figure is taken from L13.}
\label{erb}
\end{center}
\end{figure}

Finally, we can use Equation (\ref{Z1}) to predict a mean mass-metallicity relation for MS galaxies at $z \sim 2$ (assuming $\epsilon \propto (1+z)$).  This is in good qualitative agreement with the available data from Erb et al (2006), as shown in  Figure \ref{erb}.

\subsection{The stellar and dark matter content of haloes and the boost to the rsSFR} 

The product $f_{star}f_{gal}$ in this simple treatment gives the fraction of baryons that enter the halo and are transformed into stars, which, integrated up, will give $m_{star}/m_{halo}$.   If we define $\zeta$ as the mass-dependence of this quantity, $f_{star}f_{gal} \propto m_{star}^{\zeta}$, then $\zeta \sim 0.5$ is required to reconcile the faint end slope of the galaxy stellar mass function, ($\alpha_{S,star} \sim -1.45$), and the low mass end slope of the dark matter halo mass function, $\alpha_{S,halo} \sim -1.85$. From Equation (\ref{Z2}), the mean $Z(m)$ of MS galaxies gives (for constant $y_R$) the variation of $f_{star}(m)$.  Estimates of the logarithmic slope of $Z(m)$ at low masses range from $Z \propto m^{0.35}$ (Tremonti et al 2004) to $Z \propto m^{0.5}$ (Andrews \& Martini 2012, Manucci et al 2010).    This simple analysis suggests that the mass-dependence of $f_{star}$ that is implied from the mass-metallicity relation of MS galaxies (and therefore reflecting baryonic processes operating within galaxies) may be sufficient to produce the dependence of $m_{star}$ on $m_{halo}$ required to reconcile the different faint end slopes - see also Peeples \& Shankar (2012).

Secondly, as noted above, a constant $f_{star}f_{gal}$ in the regulator sets the rsSFR to be exactly the sMIR$_{DM}$.  Conversely,  if $f_{star}f_{gal}$ increases with stellar mass, then the $rsSFR$ will be boosted relative to the $sMIR_{DM}$.  Using the $\zeta$ parameter defined above, this boost to the rsSFR will be given by $rsSFR \sim (1-\zeta)^{-1} sMIR_{DM}$. It can be seen that the value of $\zeta \sim 0.5$, that is indicated from the mass-metallicity relation and from the abundance matching of mass-functions, would give a boost of the $rsSFR$ of a factor of about two above the $sMIR_{DM}$. This is about what is seen on Figure (\ref{sSFR}), especially at lower redshifts. This boost effect also explains the reversed mass-dependences of the rsSFR relative to the underlying sMIR$_{DM}$, because of the reduction in $\zeta$ associated with the flattening of the $Z(m_{star})$ relation. 

\bigskip

\noindent{\bf Acknowledgements}:   Our work has made extensive use of the COSMOS and zCOSMOS data sets, in addition to SDSS, and many of the ideas developed therein arose in discussions with collaborators in those projects, whose contributions are gratefully acknowledged.

\end{document}